\documentclass[aps,prl,twocolumn,showpacs,groupedaddress,amsmath,amssymb,floatfix,tightenlines]{revtex4-1}
\usepackage{indentfirst}
\usepackage{floatrow}
\usepackage{graphicx}
\usepackage{graphics}
\graphicspath{{figures/}}

\usepackage[T1]{fontenc}

\begin{document}
%\preprint{}


\title{Magnetic alignment of block copolymer microdomains by intrinsic chain anisotropy}
%\title{Magnetic alignment of the intrinsic anisotropy of Gaussian coils in a block copolymer}
%\title{Magnetic field induced microdomain alignment in a lamellar coil-coil block copolymer}
%\title{Magnetic field induced grain orientation of a coil-coil block copolymer}
%\title{Magnetic field directed self-assembly of a lamellar block copolymer mesophase}
%\title{Magnetic field induced alignment of a coil-coil lamellar diblock copolymer}


\author{Yekaterina Rokhlenko}
%\email{yekaterina.rokhlenko@yale.edu}
% \thanks{}
%\altaffiliation{Department of Chemical Engineering, Yale University, New Haven CT 06520}
\affiliation{Department of Chemical Engineering, Yale University,
New Haven CT 06511}
\author{Kai Zhang}
%\email{k.zhang@yale.edu}
\affiliation{Department of Mechanical Engineering and Materials Science, Yale University,
New Haven CT 06511}
\author{Steven R. Larson}
%\email[]{slarson@chem.wisc.edu}
\affiliation{Department of Materials Science and Engineering, University of Wisconsin, Madison WI 53706}
\author{Manesh Gopinadhan}
%\email[]{manesh.gopinadhan@yale.edu}
\affiliation{Department of Chemical Engineering, Yale University, New Haven CT 06511}
\author{Pawe{\l} W. Majewski}
%\email[]{pmajewski@bnl.gov}
\affiliation{Center for Functional Nanomaterials, Brookhaven National Lab, Upton NY 11973}
\author{Kevin G. Yager}
%\email[]{kyager@bnl.gov}
\affiliation{Center for Functional Nanomaterials, Brookhaven National Lab, Upton NY 11973}
\author{Padma Gopalan}
%\email[]{pgopalan@chem.wisc.edu}
\affiliation{Department of Materials Science and Engineering, University of Wisconsin, Madison WI 53706}
\author{Corey S. O'Hern}
%\email[]{corey.ohern@yale.edu}
\affiliation{Department of Mechanical Engineering and Materials Science, Yale University,
New Haven CT 06511}
\author{Chinedum O. Osuji}
%\email[]{chinedum.osuji@yale.edu}
%\homepage[]{http://www.eng.yale.edu/polymers}
%\thanks{}
\affiliation{Department of Chemical Engineering, Yale University, New Haven CT 06511}


\date{\today}

\begin{abstract}
We examine the role of intrinsic chain susceptibility anisotropy in magnetic field directed self-assembly of a block copolymer using \textit{in situ} X-ray scattering. Alignment of a lamellar mesophase is observed on cooling across the disorder-order transition with the resulting orientational order inversely proportional to the cooling rate. We discuss the origin of the susceptibility anisotropy, $\Delta\chi$, that drives alignment, and calculate its magnitude using coarse-grained molecular dynamics to sample conformations of surface-tethered chains, finding $\Delta\chi\approx 2\times10^{-8}$. From field-dependent scattering data we estimate grains of $\approx1.2$ $\mu$m are present during alignment. These results demonstrate that intrinsic anisotropy is sufficient to support strong field-induced mesophase alignment and suggest a versatile strategy for field control of orientational order in block copolymers.
\end{abstract}

 %83.80.Hj
%\keywords{}

\maketitle

%\section{Introduction}

Block copolymers (BCPs) undergo self-assembly that results in the formation of periodic structures on mesoscopic length scales. This self-assembly is driven by phase separation of chemically distinct segments, with the extent of demixing limited by their physical connectivity. The characteristic length scales therefore are largely defined by the size, or molecular weight (MW), of the polymer chain. Considerable efforts have been devoted to developing methods to reliably direct BCP self-assembly, i.e. to align BCP domains, in various device- or application-relevant geometries and length scales.

Under appropriate circumstances, magnetic fields can dictate the alignment of BCP mesophases in a highly efficacious manner. Orientational order develops subject to the presence of anisotropic field interactions that are sufficiently large to overcome thermal forces. The alignment response is a function of the anisotropy in magnetic susceptibility, $\Delta\chi=\chi_{\parallel}-\chi_{\perp}$, where the parallel direction is along the axis of highest rotational symmetry. The free energy density of the system, $\epsilon_m$, is a function of the angle $\varphi$ between the field and this parallel direction, and of the field strength, $B$, Eq. \ref{eq:field_energy_density1}. For an ensemble of anisotropic objects, $\epsilon_m$ can be expressed in terms of the scalar orientational order parameter $\langle P_2(\cos\varphi)\rangle=\langle\frac{1}{2}(3\cos^2\varphi-1)\rangle$ that describes the orientation distribution in the system, Eq. \ref{eq:field_energy_density2}. The energy difference between orthogonal alignments is $\Delta\epsilon_m=-\Delta\chi B^2/2\mu_0$. Strong alignment occurs when this magnetostatic energy for structurally coherent units, grains, is significant compared to thermal motion, that is, for $|\Delta E_m|=|\Delta\epsilon_m| V_g\gg k_BT$, where $V_g=\xi^3$ is the volume of a grain with a typical dimension $\xi$. It is therefore apparent that alignment can occur for suitably large grains, field strengths or magnetic susceptibility anisotropies.

\begin{eqnarray}
\label{eq:field_energy_density1}
\epsilon_m&=&\frac{-B^2}{2\mu_0}\left(\chi_{\parallel}\cos^2\varphi+\chi_{\perp}\sin^2\varphi\right) \\
\label{eq:field_energy_density2}
\epsilon_m&=&\frac{-\Delta\chi B^2}{3\mu_0}\langle P_2(\cos\varphi)\rangle
%\left[\frac{1}{2}\left(3\langle\cos^2\varphi\rangle-1\right)\right]
\end{eqnarray}

Prior work has relied on liquid crystalline (LC) or crystalline assembly of rigid moities integrated with the BCP to achieve a sufficiently large $\Delta\chi$ to drive alignment at reasonable field strengths. Prototypical mesogenic units such as cyanobiphenyl species have $\Delta\chi\approx 10^{-6}$ (in SI dimensionless volume units) \cite{buka1982diamagnetism,bunning1986effect}. This has been leveraged to drive alignment of several systems \cite{osuji2004alignment,tao2007hierarchical,gopinadhan2013order,deshmukh2014molecular}. $\Delta\chi$ becomes very small for typical BCPs in the absence of mesogenic groups. Shape anisotropy notwithstanding, for a lamellar diblock copolymer with volume fractions $\phi_A$ and $\phi_B$, the anisotropy, with respect to the lamellar normal, is $\Delta\chi=-(\chi_A-\chi_B)^2/[(\chi_A/\phi_A)+(\chi_B/\phi_B)]$ where $\chi_A$ and $\chi_B$ are the isotropic susceptibilities of the blocks. On this basis, for a typical symmetric non-LC BCP such as poly(styrene-b-4-vinylpyridine), $\Delta\chi\sim\mathcal{O}(10^{-10})$ \cite{sotta1999effect,LandoltBornstein_2008}. This results in a significantly smaller driving force for alignment, and as a result magnetic field alignment of such `coil-coil' BCPs has not been observed to date.

Here, we describe \textit{in situ} X-ray scattering experiments of a coil-coil diblock copolymer subjected to high magnetic fields. Cooling across the disorder-order transition in the presence of the field induces perpendicular alignment of the lamellar normals to the field direction, suggesting the presence of magnetic anisotropy of a non-trivial magnitude. The existence of this anisotropy is rationalized in terms of the  intrinsic anisotropy of individual Gaussian chains and the non-zero ensemble average of such anisotropies that is a consequence of the organization of the chain junctions along the lamellar interface between the blocks. We use molecular dynamics to simulate trajectories of representative chains and thereby estimate the intrinsic anisotropy of the system.

\begin{figure}[t]
\centering
\includegraphics[width=90mm, scale=1]{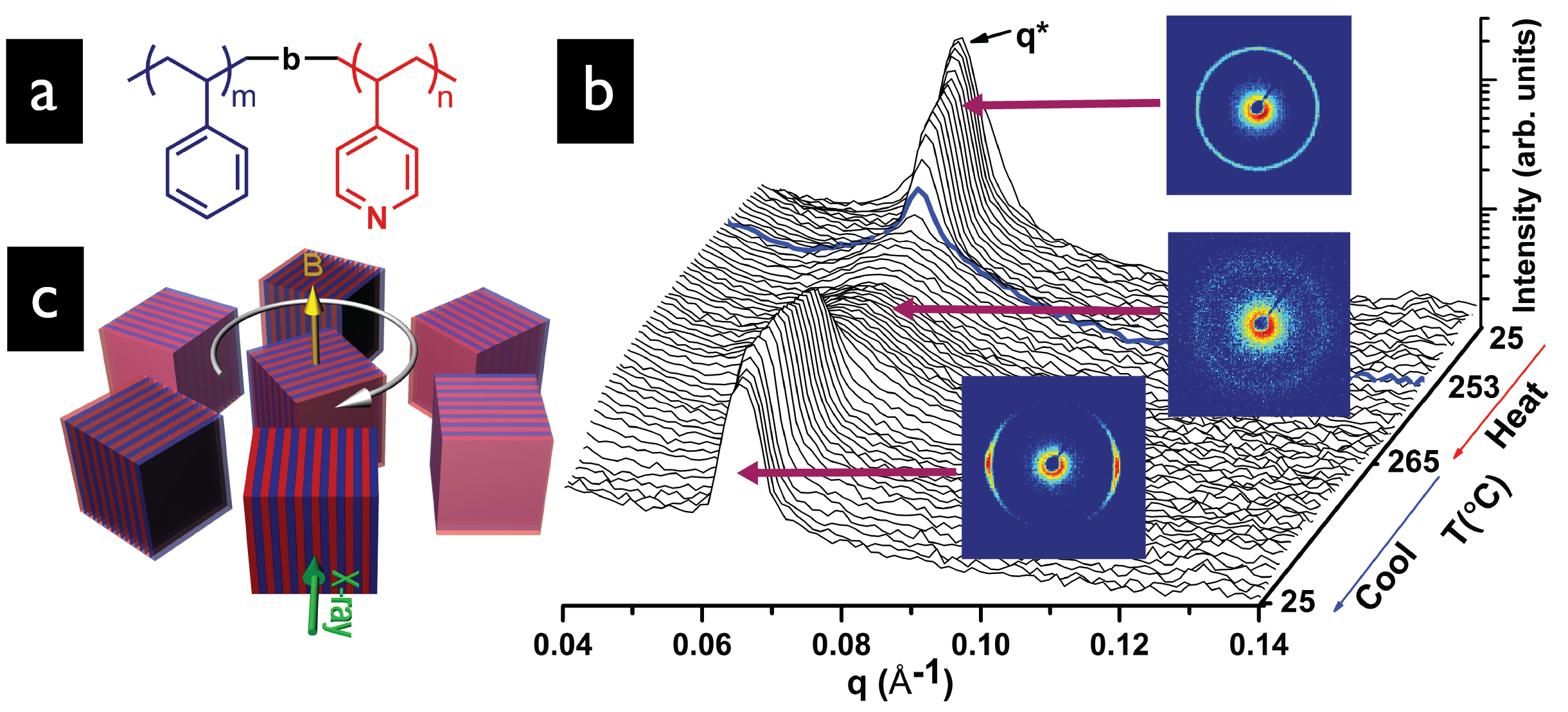}
\caption{(a) Chemical structure of PS-b-P4VP. (b) Temperature resolved SAXS data ($\mathrm{25\rightarrow 265 \rightarrow 25\, ^{\circ}C}$) under 6 T field applied vertically. 2-D diffractograms are shown inset at three representative temperatures. $T_{odt}$=253 $^{\circ}$C, d-spacing=9.5 nm. (c) Schematic illustrating degenerate system alignment with lamellar normals perpendicular to the field.}
\label{structure_saxs}
\end{figure}

The system is poly(styrene-b-4-vinyl pyridine) (PS-b-P4VP), Fig.\ref{structure_saxs}a, of MW 5.5 kg/mol (K) and PS weight fraction, $f_{PS}$=0.49, obtained from Polymer Source. All data presented here are for this 5.5K material. Consistency checks were performed using a well-purified secondary sample prepared by living anionic polymerization with MW=5.2K and $f_{PS}$=0.50. SAXS was conducted as samples were cooled at prescribed rates from 0.1 to 2 $^{\circ}$C/min across the order-disorder transition (ODT), under fields strengths from 0 to 6 T. Further details regarding experimental methods and molecular dynamics simulations are available in the Supplemental Material.

\begin{figure}[ht]
\centering
\includegraphics[width=90mm, scale=1]{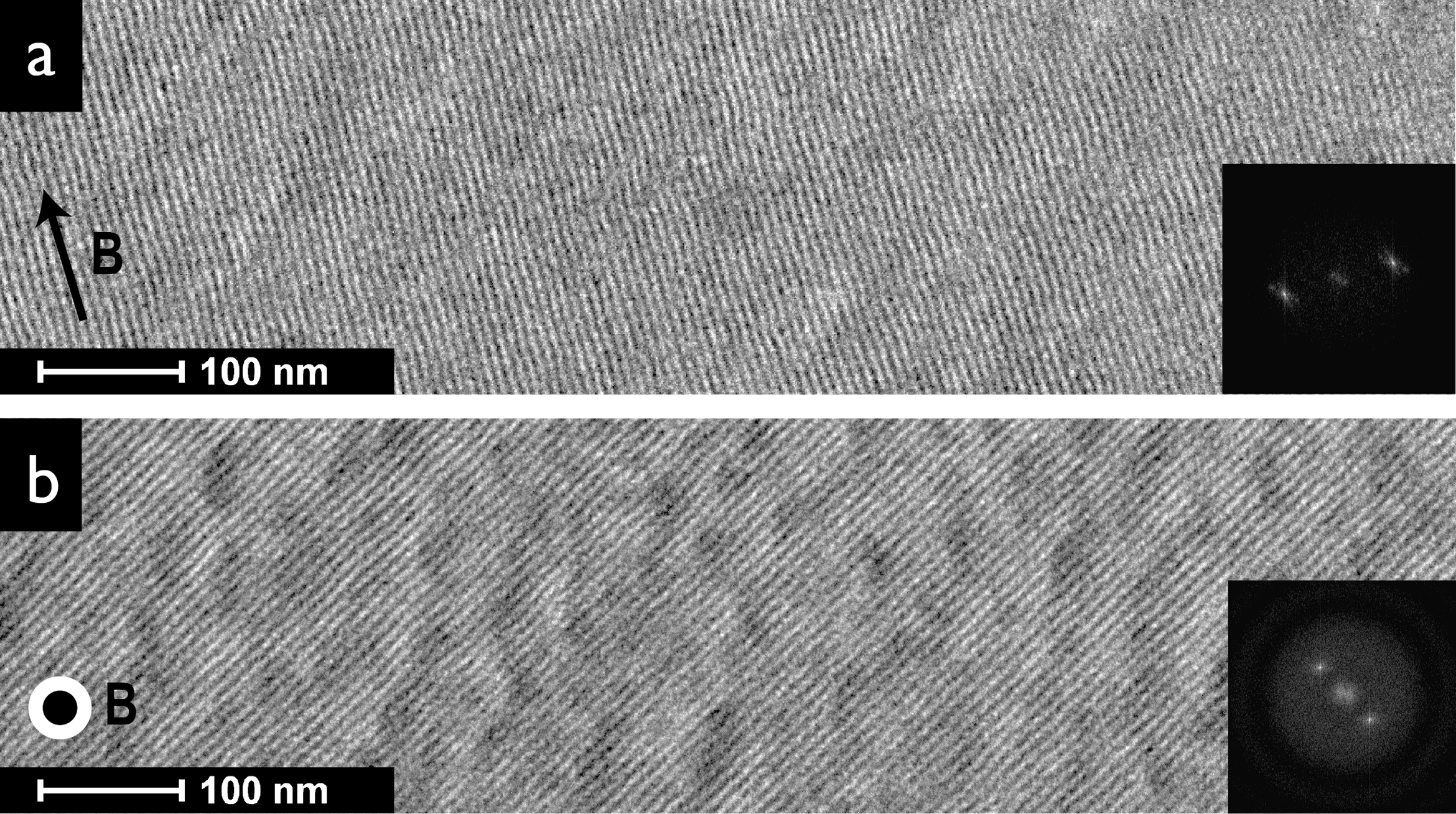}
\caption{TEM images and associated fast Fourier transforms (FFTs) of PS-b-P4VP aligned at 6 T visualized (a) perpendicular to the field direction and, (b) along the field direction.}
\label{tem_aligned}
\end{figure}

Temperature resolved SAXS data during heating and subsequent cooling at 0.3 $^{\circ}$C/min under a 6 T field is shown in Fig. \ref{structure_saxs}b. The primary peak at $q^*$=0.066 nm$^{-1}$ corresponds to the lamellar period of $2\pi/q^*$=9.5 nm and $T_{odt}$=253 $^{\circ}$C. 2-D diffractograms at select temperatures show the initial non-aligned state at room temperature, the high temperature disordered state where only correlation hole scattering is visible, and finally the aligned state produced on cooling. To the best of our ability to measure it, $\pm$ 0.5 $^{\circ}$C, $T_{odt}$ is unaffected by the field application. The concentration of intensity along the equatorial direction in the diffractogram indicates that the lamellar periodicity is perpendicular to the field which was applied vertically, i.e. along the meridional direction. The organization of lamellae with surface normals perpendicular to the applied field represents a degenerate situation, Fig. \ref{structure_saxs}c, indicating that $\Delta\chi<0$. TEM views along and perpendicular to the field, Fig. \ref{tem_aligned}, confirm the alignment of the microstructure and its degenerate nature.

\begin{figure}[ht]
\centering
\includegraphics[width=90mm, scale=1]{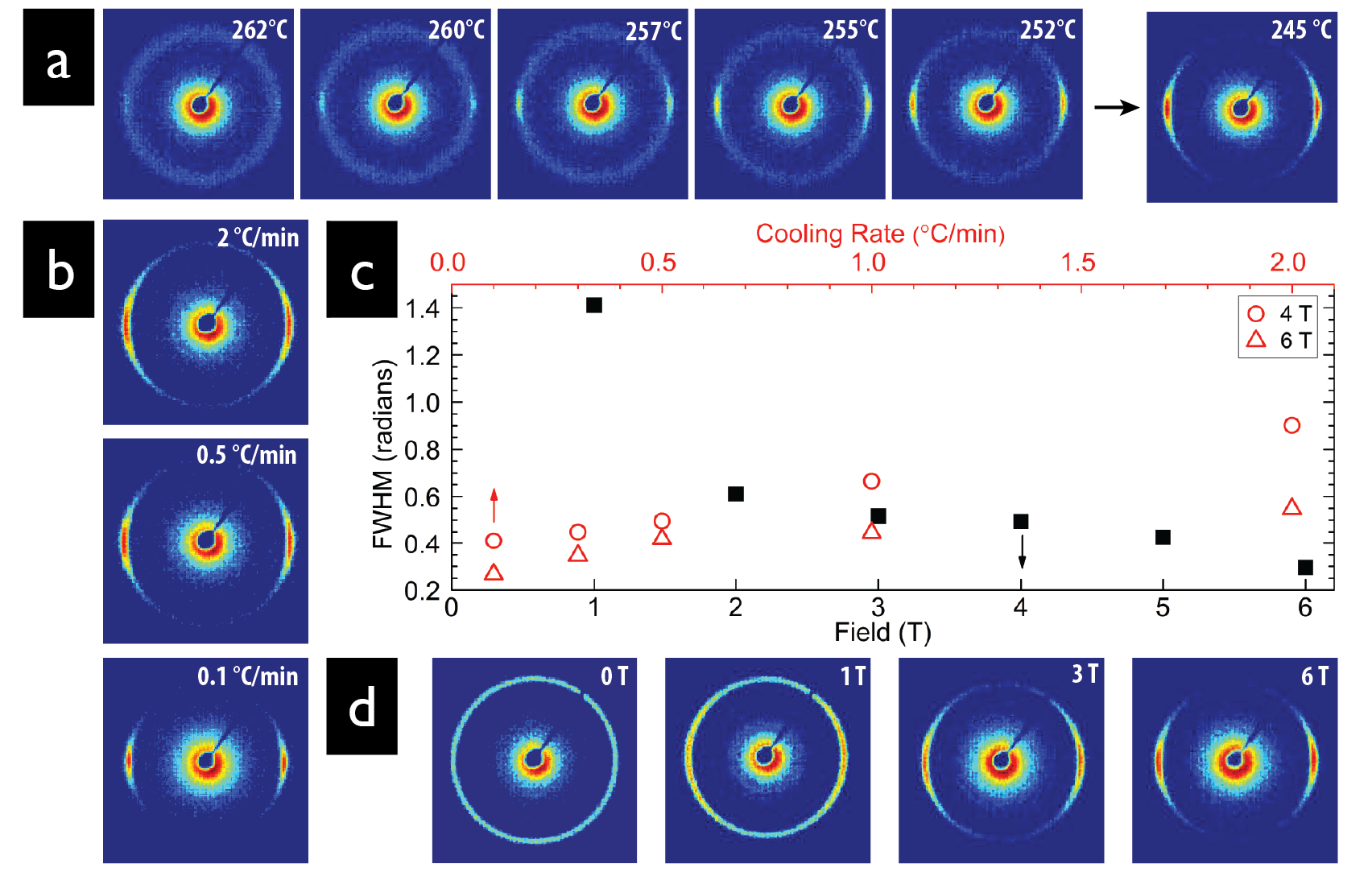}
\caption{(a) 2-D SAXS data at selected temperatures showing the emergence of aligned lamellae during cooling from the isotropic, disordered state at 0.3 $^{\circ}$C/min at 6 T. (b) Diffractograms for samples under 6 T at 245 $^{\circ}$C after cooling across $T_{odt}$ at different rates indicated. (c) Dependence of azimuthal FWHM on field strength (at 0.3 $^{\circ}$C/min) and cooling rate (at 4 and 6 T), on bottom and top x axes respectively. (d) Diffractograms at 245 $^{\circ}$C after cooling across $T_{odt}$ at 0.3 $^{\circ}$C/min at indicated field strengths.}
\label{field_rate_dependence}
\end{figure}

2-D SAXS data during passage across $T_{odt}$ are shown in Fig. \ref{field_rate_dependence}a. The evolution of order and alignment are visible in the gradual disappearance of the diffuse and azimuthally uniform intensity due to correlation hole scattering, and the concurrent emergence of Bragg scattering concentrated in arcs centered along the equatorial line. The azimuthal dependence of the scattered intensity in the Bragg peak, $I(\varphi)$, reflects the orientation distribution of the lamellar normals in the plane of the diffractogram. The full width at half maximum (FWHM) of $I(\varphi)$ depends both on the cooling rate and the field strength, with the best alignment (lowest FWHM) observed for the combination of smallest cooling rate, 0.1 $^{\circ}$C/min, and largest field, 6 T, Fig. \ref{field_rate_dependence}b-d.

The high degree of alignment suggests that the system has a magnetic anisotropy well in excess of the expected $\Delta\chi\sim\mathcal{O}(10^{-10})$. Experiments conducted with the custom synthesized secondary samples recovered similar results, dispelling any concerns regarding sample purity. Lamellar BCPs are known to form anisotropic, ellipsoidal, grains during nucleation and growth, with faster growth along the lamellar normal due to interfacial tension effects \cite{hashimoto1996nucleation,balsara2002anisotropy}. Magnetic shape anisotropy for such grains is of a trivial magnitude as the suspending fluid is a disordered melt of identical composition to the grain. Further, the shape anisotropy would lead to alignment of the long axis of the ellipsoid parallel to the field, whereas here the observed alignment is orthogonal.

The origin of the anisotropy can be understood partly by recognizing that real Gaussian chains are anisotropic entities as they possess finite end-end distances. Polarizability anisotropy has been treated in the context of freely jointed statistical segments by Kuhn \cite{Kuhn1942} yielding $\Delta\gamma=\gamma_1-\gamma_2=\frac{3}{5}(\alpha_1-\alpha_2)R^2/\langle R^2\rangle_0$ where $\alpha_1$ and $\alpha_2$ are the polarizabilities parallel and perpendicular to the bond joining statistical segments, $R^2$ is the squared end-end distance of the chain, and $\langle R^2\rangle_0$ is the unperturbed mean-squared end-end distance. The problem has also been treated in the context of the worm-like chain model by Benoit, Weill, \textit{et al.} \cite{arpin1977optical,maret1983magnetic}. While individual chains are anisotropic, the orientation of the end-end vectors in a melt are uncorrelated so there is no net anisotropy. The situation is different however for polymer brushes where one chain end is tethered to an impenetrable surface. This geometrically imposes a net anisotropy in which the orientation distribution of end-end vectors, $P_{\overrightarrow{R}}(\varphi)$, is peaked along the surface normal $\varphi=0$, and in which the width of the distribution narrows with increasing areal density of the chains. The physical analogy between surface attachment of chain ends for a brush, and chain end (junction) localization at the inter-material dividing surface of a BCP provides a segue into a treatment of chain anisotropy in BCPs. Indeed, the optical anisotropy of tethered chains has been considered in detail by Lodge and Fredrickson, for BCP melts \cite{lodge1992optical}. They note the significance of intrinsic anisotropy relative to form anisotropy in the birefringence of lamellar mesophases and highlight the importance of chain orientation due to correlation in the orientation of end-end vectors in the ordered state, relative to chain stretching.

Optical anisotropies have been investigated experimentally by depolarized Rayleigh scattering \cite{tonelli1970optical,suter1977optical,fytas1988optical} and numerically based on bond polarizability data using the rotational isomeric states approaches advanced by Nagai \cite{nagai1964photoelastic}, and by Flory and co-workers \cite{suter1977optical}. The statistical segment polarizability anisotropy $\Delta\alpha=\alpha_1-\alpha_2$ however is a quantity associated with a hypothetical construct, i.e. the statistical segment of Kuhn's freely jointed model. As such $\Delta\alpha$ cannot be linked readily to chemical structure. It has been measured however for some polymers using stress-optical coefficients \cite{Kuhn1942,stein1953determination}. These data are complemented by studies of electric field induced birefringence, the Kerr effect, in polymer solutions \cite{champion1980flow,tsvetkov1989anisotropy}. The magnetic analog in the Cotton-Mouton coefficient for magnetic birefringence provides access to the chain magnetic susceptibility anisotropy as $C_m=(\Delta n/c B^2)\sim\Delta\chi\Delta\gamma$, with $\Delta\chi$ interpreted again in the context of the chain statistical element \cite{Maret1985,tiesler1996analysis}. Precise measurements of magnetic birefringence and segmental susceptibility anisotropy of flexible chains in the melt are challenging, and there is little reported data that can be used in the present context. Additionally, attempts to link statistical segment anisotropy to that of the physical monomer lead to mixed results, particularly for polymers such as PS due to phenyl ring rotation \cite{stein1953determination}.

\begin{figure}[ht]
\centering
\includegraphics[width=75mm, scale=1]{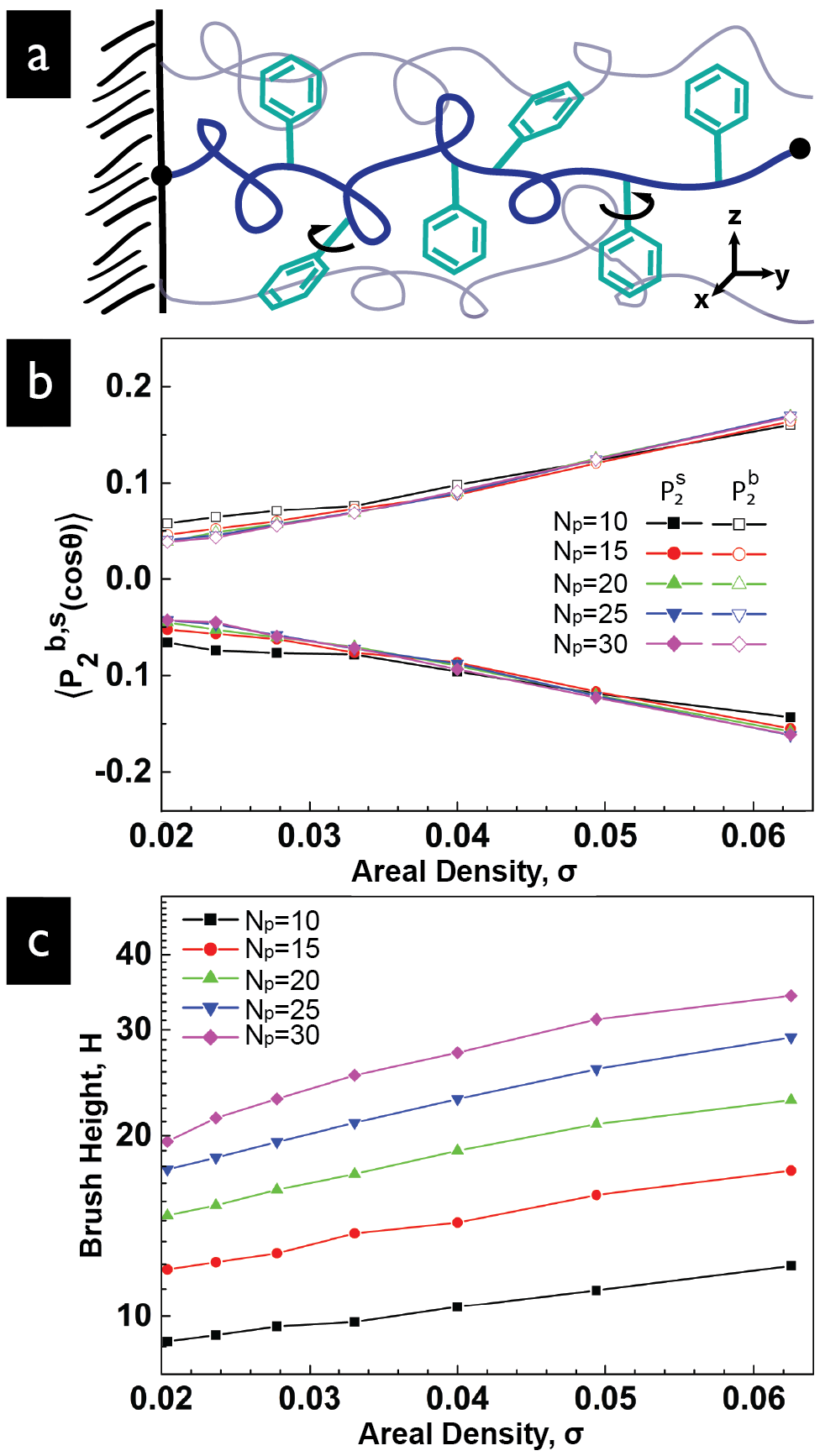}
\caption{(a) Schematic showing a single highlighted polystyrene chain tethered at an impenetrable surface. The phenyl rings can rotate around the side-bond to the backbone. (b) Simulated backbone, $\langle P_2^b\rangle$, and side chain, $\langle P_2^s\rangle$, as functions of areal density $\sigma$ for varying $N_p$. (c) Simulated brush height, $H(\sigma)$, for varying $N_p$.}
\label{simulation}
\end{figure}

In lieu of robust experimental data, we use molecular dynamics to estimate $\Delta\chi$ for PS chains tethered at an impenetrable surface. We investigate the orientational order of backbone bonds and the side bonds to the phenyl ring as a function of degree of polymerization, $N_p$, and the chain areal density at the surface, $\sigma$, Fig. \ref{simulation}. The height of the brush scales as $H\sim N\sigma^{1/2}$, as expected for chains in the melt \cite{milner1988theory}. The orientational order with respect to the surface normal, the y-axis in Fig. \ref{simulation}a, for backbone and side bonds to the phenyl ring, $\langle P_2^b(\cos\varphi)\rangle$ and $\langle P_2^s(\cos\varphi)\rangle$ respectively, increase slowly, but approximately linearly with $\sigma$. There is no statistically significant MW dependence, highlighting the lack of persistence of the chains. $\langle P_2^b\rangle$ and $\langle P_2^s\rangle$ differ in sign, reflecting the tendency of the backbone bonds therefore $\mathbf{\overrightarrow{R}}$ to lie parallel to the surface normal, and the side bonds to be orthogonal to the backbone.

The areal density of chains at the block interface is $\sigma=\rho L_0 N_A/MW$ where $L_0$ is the lamellar period. Here, $\sigma\approx 1$ nm$^{-2}$ which corresponds to $\sigma=0.033$ in simulation units. The orientational order of the phenyl ring defined by the ring normal, $\langle P_2^p(\cos\varphi)\rangle$, is also calculated (Supplemental Material). At this density, $\langle P_2^b(\cos\varphi)\rangle\approx$ 0.07 and $\langle P_2^p(\cos\varphi)\rangle\approx$ 0.04. We estimate $\Delta\chi$ for the brush based on the volume fraction, $\phi$, weighted anisotropies of the alkane backbone ($\Delta\chi^b$) \cite{shao1998magnetic} and that of the phenyl ring ($\Delta\chi^p$) \cite{LandoltBornstein_2008}. Our reference direction is the surface normal.  For the PS chains then, $\Delta\chi=\phi^b\langle P_2^b(\cos\varphi)\rangle\Delta\chi^b+\phi^p\langle P_2^p(\cos\varphi)\rangle\Delta\chi^p\approx -1.6 \times 10^{-8}$. We expect a similar contribution from the P4VP block given the near identical susceptibilities of pyridine and benzene, and so for the system overall $\Delta\chi\approx -1.6\times 10^{-8}$. (Additional details in Supplemental Material).

\begin{figure}[ht]
\centering
\includegraphics[width=75mm, scale=1]{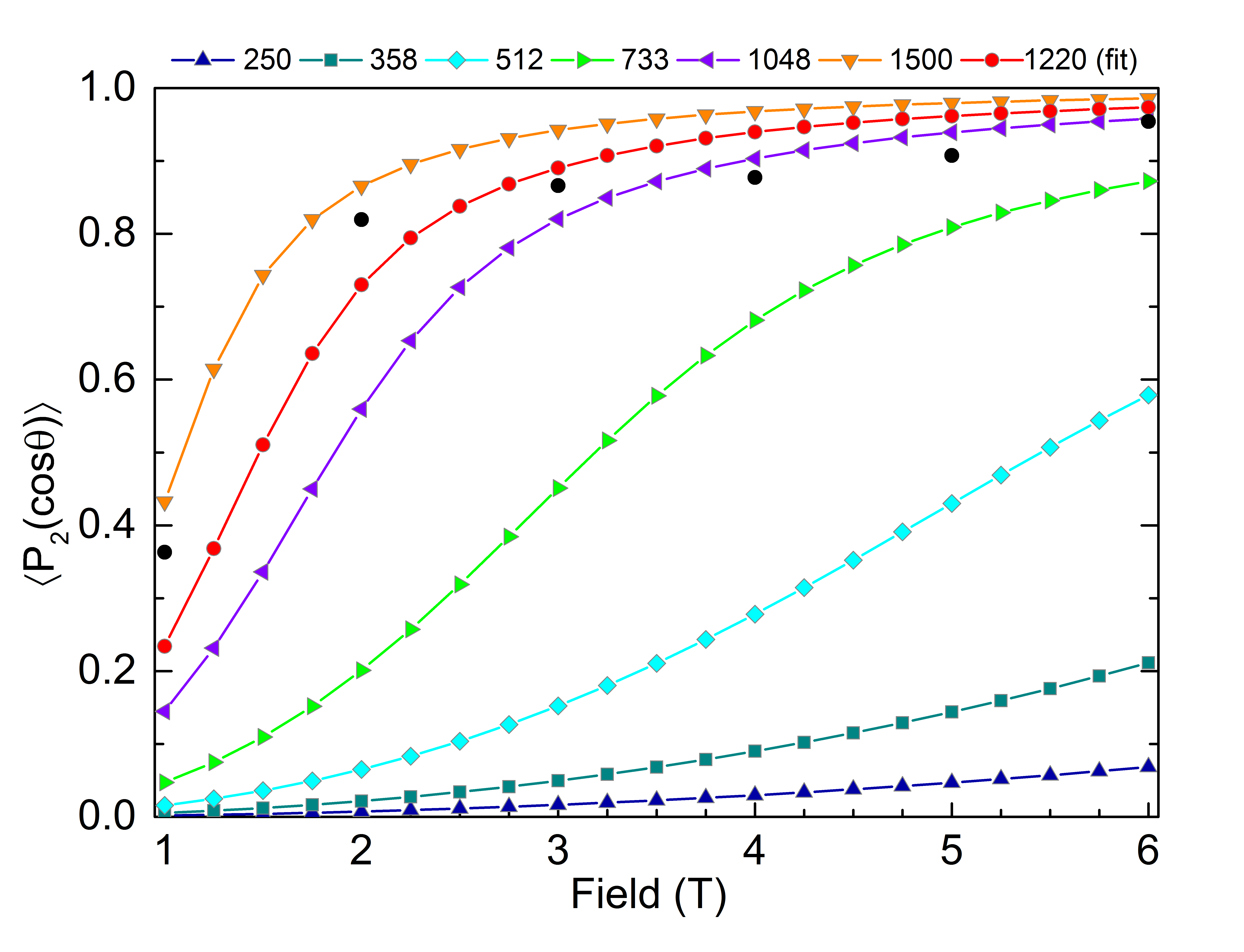}
\caption{Field-dependent orientational order parameters (Eq. \ref{eq:P2_energy2}) for various grain sizes, $\xi$, for $\Delta\chi= -1.6\times 10^{-8}$ at 253 $^{\circ}$C. Black circles show experimentally derived values for samples cooled 0.3 $^{\circ}$C/min. Fitting Eq. \ref{eq:P2_energy2} (red circle trace) to experimental data yields $\xi\approx 1.2$ $\mu$m.}
\label{grain_size_fit}
\end{figure}

We use the estimated $\Delta\chi$ to consider the alignment data of Fig. \ref{field_rate_dependence}. The azimuthal intensity dependence $I(\varphi)$ reflects the probability of observing lamellar normals at a given angle $\varphi$ with respect to the field direction. This probability $p(\varphi,B)$, is governed by a Boltzmann factor incorporating the  angle-dependent magnetostatic energy, $E_m(\varphi,B)=-(B^2/2\mu_0)\Delta\chi\xi^3\cos^2\varphi$, for a characteristic grain size, $\xi$, Eq. \ref{eq:P2_energy1}. The orientational order parameter resulting from this probability distribution may be evaluated by integration, Eq. \ref {eq:P2_energy2}. The orientational order parameters are obtained as a function of field strength from Gaussian fits of the SAXS intensity profiles. We estimate the representative grain size by fitting the experimentally determined $\langle P_2\rangle$ to those calculated using Eq. \ref{eq:P2_energy2}, yielding $\xi\approx 1.2\ \mu$m. This corresponds to a field interaction $\Delta E_m>10^2$ kT. The results are shown in Fig. \ref{grain_size_fit}, with calculated field dependent $P_2$ for different grain sizes. It is important to note that the estimated grain size only provides the characteristic dimensions of a structural unit which would, at steady state, reproduce the orientation distribution measured in the aligned samples at room temperature. Kinetic effects related to changes in temperature dependent nucleation and growth rates, as well as any Ostwald ripening, during the cooling ramp do not factor into this analysis.
 \begin{eqnarray}
%\label{eq:P2_energy1}
 % E_m(\varphi,B)&=&-(B^2/2\mu_0)\Delta\chi\xi^3\cos^2\varphi \\
\label{eq:P2_energy1}
 p(\varphi,B)&=&\frac{e^{-E_m/kT}\sin\varphi\,d\varphi}{\int^{\pi}_0e^{-E_m/kT}\sin\varphi\,d\varphi} \\
 \label{eq:P2_energy2}
\langle P_2(\cos\varphi)\rangle&=&\frac{\int^{\pi}_0 \frac{1}{2}(3\cos^2\varphi-1)\,e^{-E_m/kT}\,\sin\varphi\,d\varphi}{\int^{\pi}_0e^{-E_m/kT}\,\sin\varphi\,d\varphi}
\end{eqnarray}

The grain size determined by fitting is in modest agreement with an estimated size of $\approx 3$ $\mu$m from sampling TEM images of field aligned samples. To provide better quantification we independently measured the grain size using a recently developed `variance scattering' technique \cite{yager2014metrics}. This approach statistically analyses the azimuthal intensity variation of a primary Bragg peak, from which the number of independent scatterers, and thus the grain size, can be quantified. On this basis we estimate $\xi\approx 5-6\, \mu$m at room temperature (further details in Supplemental Material). The fact that TEM and scattering derived grain sizes exceed the estimate based on the orientation distribution reflects the occurrence of microstructure coarsening without accompanying improvement in the alignment under the field. We should therefore view the estimated $\xi$ as an upper bound for the characteristic grain size that pertained during the period in which the system was responsive to the field.

BCP nucleation and growth kinetics have been explored experimentally and theoretically with some success \cite{fredrickson1989kinetics,Balsara1996,hashimoto1996nucleation,goveas1997dynamics,chastek2004grain}. The slow dynamics of high MW BCPs has largely restricted studies to unentangled melts and concentrated solutions as a matter of convenience. Grain sizes of $\sim 0.5-5$ $\mu$m are not uncommon. It is therefore reasonable to expect that sufficiently large grains can be prepared in a variety of  systems to permit control of orientation using magnetic fields. The slow dynamics of entangled melts and the inaccessibility of the ODT in high MW systems suggest that this strategy would be limited to melts below entanglement MW and concentrated solutions. It is clear that the aromatic nature of PS-b-P4VP contributes to a markedly larger $\Delta\chi$ than one would encounter for other common but non-aromatic polymers such as polyethylene oxide or polymethylmethacrylate. The fact however that $E_m\sim\xi^3$ means that an order of magnitude decrease in $\Delta\chi$ can be compensated by an increase in $\xi$ by a factor of just over 2, for the purposes of alignment.

\begin{acknowledgments}
This work was supported by NSF under DMR-1119826 and DMR-1410568. Facilities use was supported by YINQE. Additionally, this research used resources of the Center for Functional Nanomaterials, which is a U.S. DOE Office of Science Facility, at Brookhaven National Laboratory under Contract No. DE-SC0012704. The authors thank Nitash Balsara and Zhen-Gang Wang for fruitful discussions, and Mike Degen (Rigaku Inc.) and AMI Inc. for technical support.
\end{acknowledgments}

\bibliographystyle{apsrev}
\bibliography{nonLC_alignment}

\section{SUPPLEMENTAL MATERIAL}

\newcommand{\beginsupplement}{%
        \setcounter{table}{0}
        \renewcommand{\thetable}{S\arabic{table}}%
        \setcounter{figure}{0}
        \renewcommand{\thefigure}{S\arabic{figure}}%
}

%%%%%%%%%%%%%%%%% END OF PREAMBLE %%%%%%%%%%%%%%%%

%\beginsupplement

%\baselineskip24pt


%\maketitle

\section*{Sample Preparation}

Poly(styrene-b-4-vinylpyridine) (PS-b-P4VP) with molecular weight MW=5.5 kg/mol (2.7K PS/2.8K P4VP) and polydispersity index \DJ$\approx$ 1.2 was obtained from Polymer Source and used as received. A secondary sample with MW=5.2K (2.6K PS/2.6K P4VP) was prepared by living anionic polymerization. Material was pressed into an aluminum sample stage supported by thin Kapton windows at 160$^{\circ}$C to form cylindrical discs a few mm thick, and 1 cm in diameter. Small-angle x-ray scattering (SAXS) was performed in vacuum with a pinhole-collimated instrument (Rigaku SMAX3000) using Cu K$\alpha$ radiation ($\lambda= 1.542$ {\AA}). The beam has a 1 mm diameter at the sample plane and the accessible range of scattering vectors is 0.02 to 0.35 {\AA}$^{-1}$. SAXS data were calibrated using a silver behenate standard. 2-D patterns were integrated azimuthally to provide 1-D representations of I(q), where $q = (4\pi/\theta)\sin\theta$ and 2$\theta$ is the scattering angle. Temperature-resolved measurements were performed with a custom made hot-stage with an associated Omega temperature controller. Samples were loaded onto the temperature controlled aluminum stage within a superconducting magnet (American Magnetics Inc.) with a flux density between 0 and 6 T. Samples were subjected to varying cooling rates and field strengths. typically by heating to 265$^{\circ}$C and then cooling to room temperature in the presence of the field. For temperature resolved SAXS measurements, data were acquired in a quasi-isothermal experiment in which the sample was held at constant temperature for several minutes at closely spaced temperature points. For TEM, $\sim$50 nm thin sections were prepared using a diamond knife mounted on a Leica EM UC7 ultramicrotome. Sections were collected on TEM grids after flotation onto water and stained inside an iodine chamber for 1 hour prior to imaging. Sectioned samples were then visualized by FEI Tecnai Osiris TEM with an accelerating voltage of 200 kV.

\begin{figure}[h!]
  \centering
    \includegraphics[width=85mm]{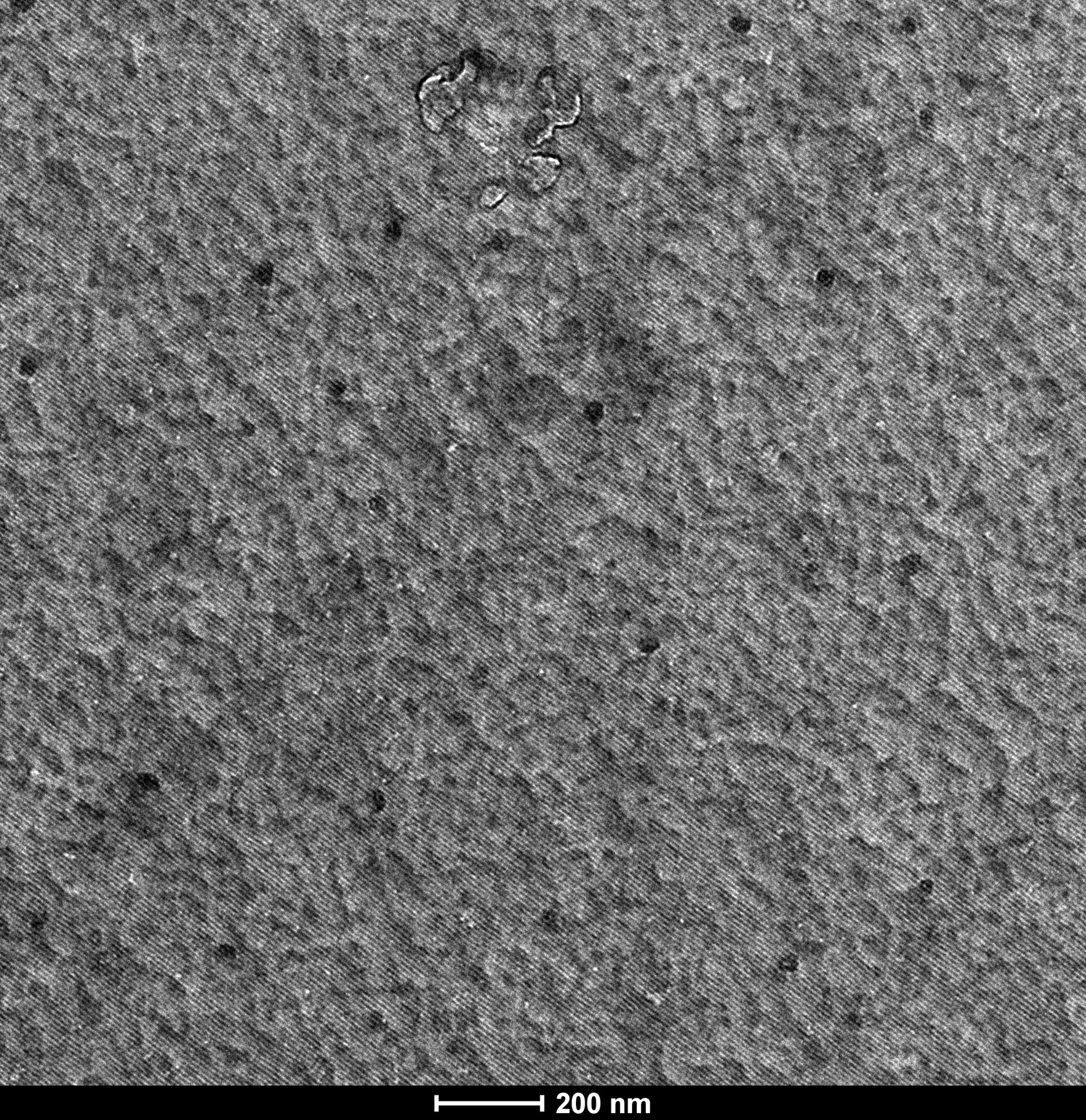}
    \caption{TEM image of PS-b-P4VP (2.7k/2.8k) prepared as described above but cooled in the absence of the field at 0.3 $^{\circ}$C/min. Here, we show the lamellar microstructure (9.5 nm periodicity) of a representative single grain, $>$2 $\mu$m $\times$ 2 $\mu$m. Grains of this size and larger are present throughout the sample whether or not a magnetic field is applied. }
\end{figure}

%\clearpage
\section*{ODT Determination}
To calculate the order disorder temperature of PS-b-P4VP (2.7k/2.8k) we plot inverse intensity of the primary scattering peak, $q^{\ast}=\ 0.066$ nm$\mathrm{^{-1}}$ vs. inverse temperature, in $K$. $T_{odt}$ is extracted as the value of temperature at $I^{-1}=0$ by extrapolating a linear fit of the inverse intensity drop during ordering.

{\centering
Fitted data:\hspace{15 pt}
$1/I=(-18445.795/T)+35.066$
$\Rightarrow T_{odt}=252.9 ^{\circ}$C

\begin{figure}[h!]
  \centering
    \includegraphics[width=90mm]{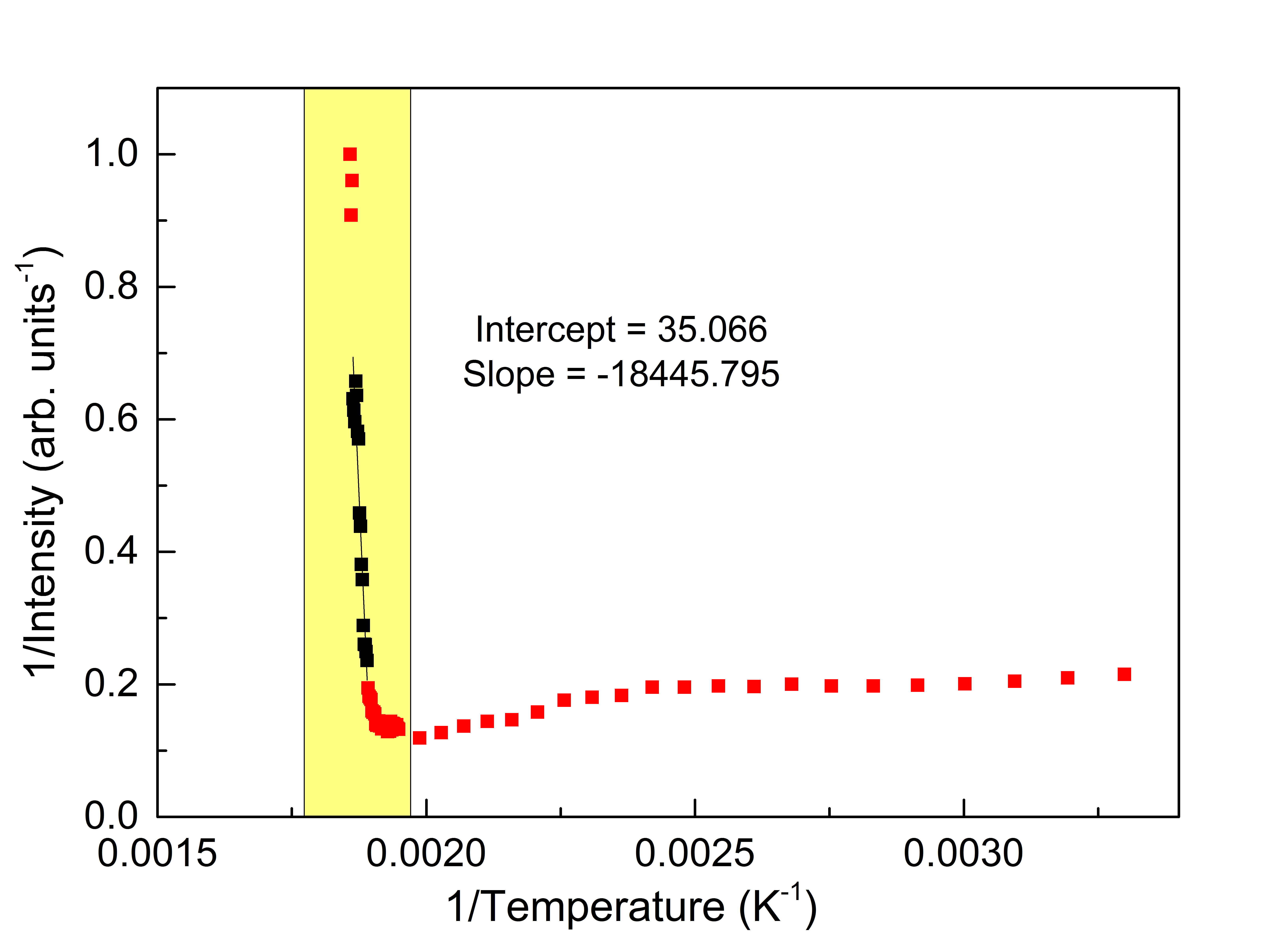}
    \caption{Plot of inverse scattered intensity vs. inverse temperature used for $T_{odt}$ determination.}
\end{figure}
}

\section*{Grain size determination by variance analysis}

For this measurement and calculation, PS-b-P4VP (2.7k/2.8k) was cooled 0.3$^{\circ}$C/min in the absence of the field in the same set up as described above. Kapton windows were gently removed, and the sample was mounted on a tripod polisher using a hot glue ($T_m=80^{\circ}$C) and wet-polished using a descending sequence of diamond lapping films on a rotary disc polisher. Before polishing the other side of each sample to its final thickness (100-200 $\mu m$), the specimens were glued onto a  25 $\mu m$ thick Kapton foil. The procedure yielded SAXS specimens with smooth and parallel sides of known thickness $(t)$ determined by a micrometer down to a $\pm$ 2$\mu m$ accuracy. The actual thickness was recorded for each sample to be further used in calculation of the scattering volume.

Small angle scattering experiments were performed on a Bruker Nanostar system fitted with a 2048 by 2048 pixel Vantex-2000 area detector positioned $\sim$ 100 $cm$ from the sample. The system was configured to high resolution mode using 300 $\mu m$ scatterless Ge pinholes. The beam profile was assessed by measuring transmitted intensity during two orthogonal scans of a metal edge located at the sample position across the footprint of the beam. The vertical and horizontal profiles of the beam were reconstructed by differentiating the recorded data and fitting them to a Gaussian function. The horizontal and vertical full width at half maximum (FWHM) parameters were used as a measure of the beam width and height, respectively (both slightly larger than $300 \mu m$). The scattering volume is calculated as $V=FWHM_h \times FWHM_v \times t$. SAXS data was collected under vacuum, in transmission mode, with each frame collected from an independent spot for 1 h. Silver behenate was used as the q-space calibrant. After the measurements, the patterns were spacially unwarped using vendor software, and exported for further analysis using custom-written Python routines.

Two-dimensional diffractograms were azimuthally integrated to yield I vs.q curves. The intensity profile of the first order 001 ($q*$) peak was fitted to a Gaussian function to measure its position q, and characterize breadth in radial direction, $\sigma_q$. From this we computed the expected peak width in the azimuthal direction $\chi$ (i.e., along the arc of the scattering ring), $\sigma_{\chi}$. The intensity along the 001 ring in the angular $(\chi)$ direction was extracted by integrating the full peak width (in the q direction), and subtracting the local background scattering just outside the peak. This I vs.$\chi$  data was smoothed using a Gaussian function with a width, $\sigma_{\chi}$. This averages out noise in the signal (e.g., shot-noise due to counting statistics) without averaging out the intensity variation arising from the grain structure. We then quantitatively convert this intensity variation into an estimate for the numbder of independant grains within the scattering volume using a previously published method \cite{yager2014metrics}.

Briefly, this method involved computing the relative standard deviation of I vs $\chi$, which can be converted into an estimate of the number of grains ($N_g$) using knowledge of the peak width ($\sigma_{\chi}$), and the peak multiplicity (m=2 for lamellar peaks). This analysis implicitly assumes an isotropic distribution of grain orientations, with relatively well-defined grain boundaries. The average grain size can then simply be computed from the known scattering volume (V) as $\xi=(3V/4\pi N_g)^{1/3}$, which assumes roughly spherical grains. We minimize the error of this estimate by combining the results from different sample thicknesses.

%\newpage

\section*{Molecular Dynamics Simulations}

We perform molecular dynamics (MD) simulations on an array of $N=5\times5=25$ coarse grained polystyrene chains with degree of polymerization $N_p=10,15,20,25,30$ sitting on a square lattice of spacing, $d$. Each monomer (styrene) has $n=8$ coarse-grained atoms including two $sp^3$ backbone carbons and six $sp^3$ phenyl ring carbons. The total potential energy as a function of atom positions $(\bf{r_1,r_2,...r_N})$ is defined as:

%\begin{multline}
\begin{widetext}
\begin{equation}
U{(\bf{r_1,r_2,...r_N})}=\sum\limits_{i<j}u_{LJ}(r_{ij}) + \sum\limits_{<ij>}u_{spring}(r_{ij}) + \sum\limits_{<ijk>}u_{angle}{(\bf{r_i,r_j,r_k})} + \\ \sum\limits_{<ijkl>}u_{torsion}{(\bf{r_i,r_j,r_k,r_l})} + \sum\limits_iu_{wall}(\bf{r_i})
\end{equation}
\end{widetext}
%\end{multline}

\noindent The first term summing over all pairs of atoms is the Lennard-Jones (LJ) potential,
\begin{equation}
u_{LJ}(r_{ij})=4\epsilon\left[(\frac{\sigma_0}{r_{ij}})^{12}-(\frac{\sigma_0}{r_{ij}})^6\right],
\end{equation}

\noindent where $\epsilon$ and $\sigma_0$ sets the unit of energy and length of the system respectively.

The second term, summing over all pairs of atoms that are connected by chemical bonds, is the harmonic spring potential,
\begin{equation}
u_{spring}(r_{ij})=\frac{1}{2} k_s (r_{ij}-r_{0})^2,
\end{equation}

where the equilibrium bond length is $r_0=\sigma_0$, and the bond stiffness $k_s=9000$ and $18000$ $\epsilon$/$\sigma_0^2$ for single and double carbon-carbon bonds, respectively.

The third term, summing over every three atoms that form a bond angle $\theta$, is
\begin{equation}
u_{angle}{\bf({r_i,r_j,r_k})}=u_{angle}(\theta)=\frac{1}{2}k_a(\theta-\theta_0)^2,
\end{equation}

\noindent where the equilibrium bond angle $\theta_0=109.5^{\circ}$ for $sp^3$ carbons and $120^{\circ}$ for $sp^2$ carbons. The bond angle stiffness is $k_s=400\ \epsilon$/rad$^2$ for $sp^3$ carbons and $18000\ \epsilon$/rad$^2$ for $sp^2$ carbons. This potential keeps the phenyl ring rigid and flat.

The fourth term, summing over every four consecutive backbone atoms or carbon atoms connected by $sp^3$-$sp^2$ side chains that form a dihedral torsion angle $\phi$, is

\begin{widetext}
\begin{equation}
u_{torsion}{\bf({r_i,r_j,r_k,r_l})}=u_{torsion}(\phi)=c_3 cos^3(\phi-\phi_0)-c_1 cos(\phi-\phi_0)+c_3-c_1
\end{equation}
\end{widetext}

where $c_1,c_3=6,10\ \epsilon$ and the phase shift $\phi_0=180{^\circ}$ such that the four atoms tend to stay in the $trans$ configuration.

The last term describes the purely repulsive LJ wall at $y=0$ where polymer chains are attached. In the simulation, the first atom of each chain is pinned at $y=0$. Periodic boundary conditions are applied in the x and z directions.

All simulations are run at the reduced temperature $T=3.0\  \epsilon/k_B$ controlled by the Gaussian constraint thermostat, where $k_B$ is Boltzmann's constant. Polymer chains are initialized as straight lines with side chains perpendicular to the backbone. We first equilibrate the system for $10^5$ time steps until the energy fluctuations become stable, and then sample thermodynamic quantities over $10^6$ time steps with a time step $dt=0.001 \sqrt{m\sigma^2/\epsilon}$.

We measure the backbone $P_2^b$, side chain $P_2^s$, and phenyl ring $P_2^r$ bond order parameters with respect to the director $\hat{y}$, normal to the surface, where $\theta_k$ is the angle between the bond or ring normal, and the director $\hat{y}$.

\begin{equation}
\langle P_2^{b,s,r}\rangle=\left\langle \frac{1}{N_{b,s,r}}\sum\limits_{k=1}^{N_{b,s,r}}\left(\frac{3}{2}cos^2\theta_k-\frac{1}{2}\right)\right\rangle
\end{equation}

\noindent There are $N_b=(2N_p-1)/N$ backbone vectors, $N_s=N_pN$ side chain vectors, and $N_r=N_pN$ phenyl ring normal vectors. For comparison, we also measure $P_2^r$ with respect to the other two axes $\hat{x}$ and $\hat{z}$ parallel to the surface.

The average height $H$ of polymer brushes as a function of their surface density $\sigma=1/d^2$ is measured for lattice spacing $d=4.0,4.5,5.0,5.5,6.0,6.5,7.0\sigma_0$. $H$ is defined as the average of the maximum height coordinates of individual chains. The scaling exponent $\alpha$ in $H\sim(1/d^2)^{\alpha}$ is extracted as $\frac{1}{2}$ as expected for chains in the melt.

\section*{Estimation of susceptibility anisotropy, $\Delta\chi$}

\begin{figure}[!h]
           %\begin{floatrow}
             \ffigbox{\includegraphics[width=70mm]{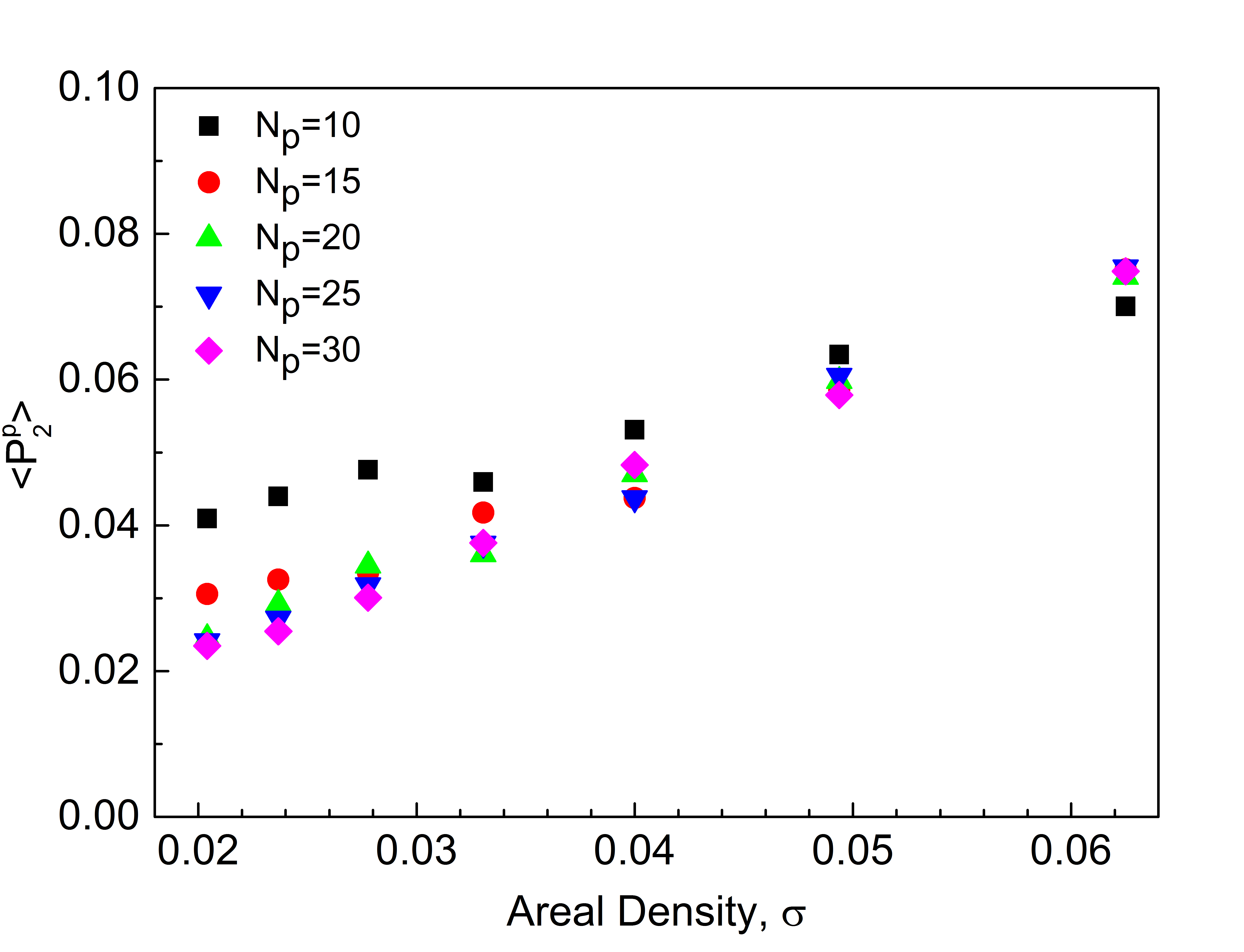}}{\caption{$\langle P_2^p \rangle$, phenyl ring order parameter, as calculated in MD simulation, plotted against areal density of chains, $\sigma$.} \label{case1}}
             \ffigbox{\includegraphics[width=70mm]{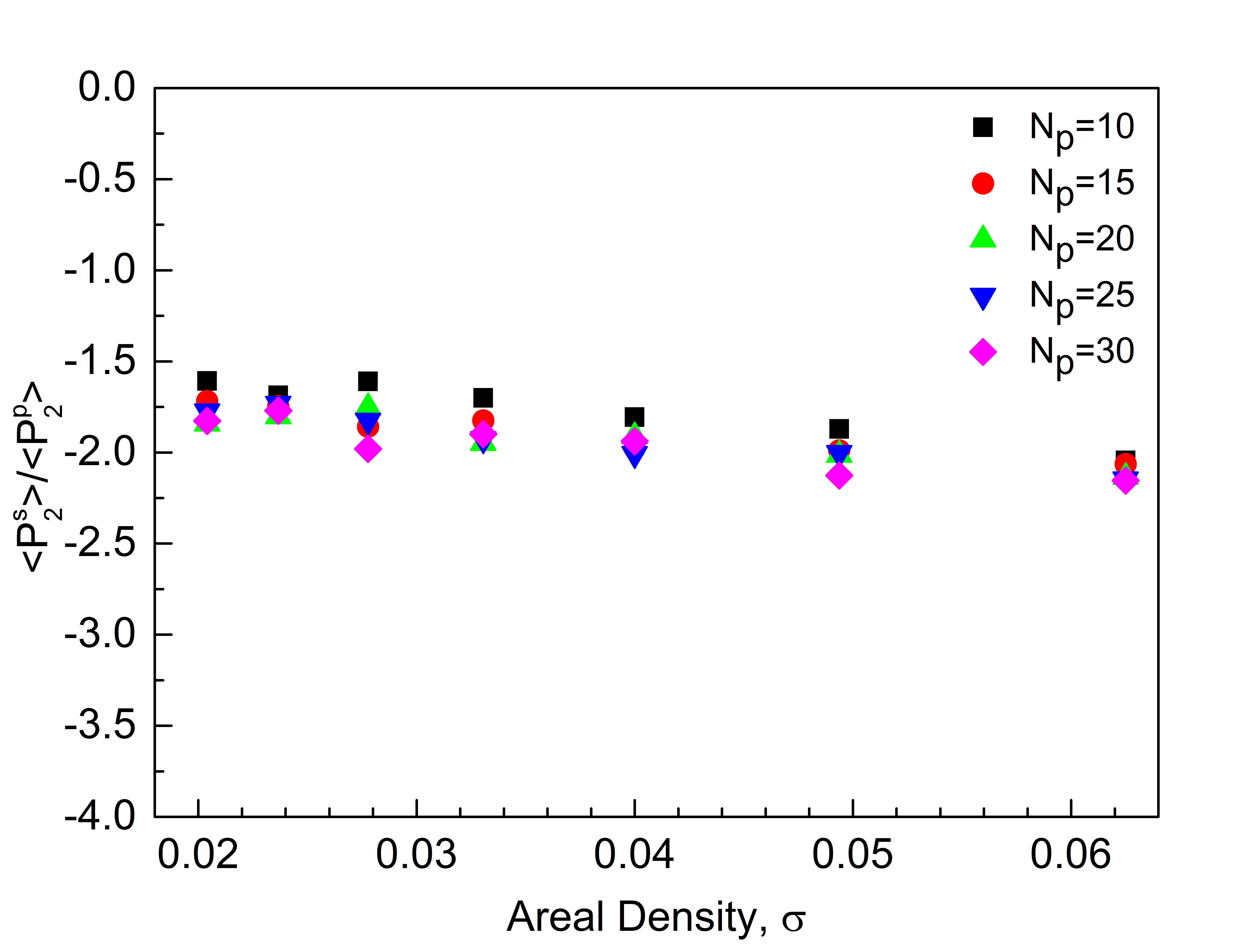}}{\caption{Ratio of side chain order parameter, $\langle P_2^s \rangle$ to phenyl ring order parameter, $\langle P_2^p \rangle$, plotted against areal density of chains, $\sigma$. }\label{zoom}}
           %\end{floatrow}
\end{figure}

As expected, there is a geometrical relationship between the orientational order of the side chain, $\langle P_2^s \rangle$, and the orientational order of the phenyl ring, $\langle P_2^p \rangle$, Figs. \ref{case1} and \ref{zoom}. The calculated ratio of the two has a magnitude of $\approx$ 1.8 which compares well with the expected value, 2.

\noindent We use the simulated orientational order parameters of the backbone and phenyl ring, $\langle P_2^b \rangle$ and $\langle P_2^p \rangle$, to estimate the overall anisotropy, $\Delta\chi$, present in the system. Although the anisotropy of benzene is commonly considered to be positive as it is viewed intuitively with respect to the plane of the ring, here we utilize the formal specification which is with respect to the axis of highest rotational symmetry, the ring normal, and thus $\Delta\chi^b<0$.

\begin{widetext}
\begin{equation}
\Delta\chi=\phi^b\langle P_2^b(cos\theta)\rangle\Delta\chi^b+\phi^p\langle P_2^p(cos\theta)\rangle\Delta\chi^p \approx -1.6 \times 10^{-8}
\end{equation}
\end{widetext}

{\small
\indent Anisotropy of alkane backbone, $\Delta\chi^b\approx -5 \times 10^{-8}$\\
\indent Anisotropy of phenyl ring, $\Delta\chi^p\approx -5 \times 10^{-7}$\\
\indent Orientational order of backbone, $\langle P_2^b(cos\theta)\rangle\approx 0.07$\\
\indent Orientational order of phenyl ring, $\langle P_2^p(cos\theta)\rangle\approx 0.04$\\
\indent Phenyl ring volume fraction (estimated as weight fraction), $\mathrm{\phi_p\approx (M^{phenyl}/M^{styrene})\approx 0.75}$.\\
\indent Backbone volume fraction (estimated as weight fraction), $\phi_b\approx1-\phi_p\approx 0.25$
}

%\newpage
%\clearpage

\end{document}